# Universal quantum phase transition from superconducting to insulating-like states in pressurized $Bi_2Sr_2CaCu_2O_{8+\delta}$ superconductors

Yazhou Zhou[1]*, Jing Guo[1,6]*, Shu Cai[1]*, Jinyu Zhao[1,4], Genda Gu[2], Chengtian Lin[3], Hongtao Yan[1,4], Cheng Huang[1,4], Chongli Yang[1], Sijin Long[1,4], Yu Gong[5], Yanchun Li[5], Xiaodong Li[5], Qi Wu[1], Jiangping Hu[1,4], Xingjiang Zhou[1,4,6], Tao Xiang[1,4] and Liling Sun[1,4,6]†

[1]Institute of Physics, National Laboratory for Condensed Matter Physics, Chinese Academy of Sciences, Beijing, 100190, China

[2] Condensed Matter Physics & Materials Science Department, Brookhaven National Laboratory, NY, 11973-5000 USA

[3]Max-Planck-Institut fu¨r Festko¨rperforschung, D-70569 Stuttgart, Germany

[4]University of Chinese Academy of Sciences, Department of Physics, Beijing 100190, China

[5]Institute of High Energy Physics, Chinese Academy of Science, Beijing 100049, China

[6]Songshan Lake Materials Laboratory, Dongguan, Guangdong 523808, China

Copper oxide superconductors have been continually fascinating the communities of condensed matter physics and material sciences because they host the highest ambient-pressure superconducting transition temperature ($T_c$) and mysterious physics [1-3]. Searching for the universal correlation between the superconducting state and its normal state or neighboring ground state is believed to be an effective way for finding clues to elucidate the underlying mechanism of the superconductivity. One of the common pictures for the copper oxide superconductors is that a well-behaved metallic phase will present after the superconductivity is entirely suppressed by chemical doping [4-8] or application of the magnetic field [9]. Here, we report the first observation of universal quantum phase transition from superconducting state to insulating-like state under pressure in the under-, optimally- and over-doped $Bi_2Sr_2CaCu_2O_{8+\delta}$ (Bi2212) superconductors with two $CuO_2$ planes in a unit cell. The same phenomenon has also been found in the $Bi_2Sr_{1.63}La_{0.37}CuO_{6+\delta}$ (Bi2201) superconductor with one $CuO_2$ plane and the $Bi_{2.1}Sr_{1.9}Ca_2Cu_3O_{10+\delta}$ (Bi2223) superconductor with three $CuO_2$ planes in a unit cell. These results not only provide fresh information on the cuprate superconductors but also pose a new challenge for achieving unified understandings on the mechanism of the high-$T_c$ superconductivity.

Although a huge body of experimental investigations has been made for the copper oxide (cuprate) superconductors since they were discovered for more than thirty years [10,11], the correlation between the superconducting state and its normal state or the neighboring ground state is widely debated [2,6,12-14]. By changing the chemical makeup of interleaved charge-reservoir layers, electrons can be added to or removed from the $CuO_2$ planes, resulting in the suppression of the antiferromagnetic insulating state of the parent compound [2]. As the doping level reaches a critical one, superconductivity presents and its transition temperature ($T_c$) grows to a maximum upon doping to an optimal one, then declines for higher doping, and finally vanishes at a maximum doping level [2,5,7,9,15]. $T_{c\text{-max}}$ are referred to as optimal-doped ones. It is important to recognize that once the superconducting state is completely suppressed by the chemical doping, the material undergoes a quantum phase transition from a superconducting state to a metallic state [16-18]. However, the detailed experimental studies on the breakdown of the quantum state in cuprates are still lacking, which may be crucial for understanding how the superconducting state melts into or emerges from its neighboring ground states.

Pressure is an alternative method of tuning superconductivity beyond the chemical doping or external magnetic field, and it can provide significant information on the evolution among superconductivity, electronic state, and crystal structure without changing the chemical composition. On the other hand, it can also provide valuable assistance in the search for the superconductors with higher values of $T_c$ at ambient pressure by the substitution of the smaller ions [19]. A notably successful application of this strategy leads to the discoveries of the important cuprate and Fe-based superconductors [11,20,21]. Therefore, high-pressure studies on superconductivity can benefit not only for searching new superconductors but also for deeper understandings on the correlation between the superconducting and its neighbor normal or ground states [22-26]. To reveal how the superconducting state or non-superconducting state develops, a central issue for understanding the high-$T_c$ superconductivity in cuprates, we performed a series of high-pressure investigations by employing our newly developed state-of-the-art technique, a combined *in-situ* high-pressure measurements of the resistance and alternating current (*ac*) susceptibility for the same sample at the same pressure. The studied samples that have been investigated broadly by variety of methods [26-31] are the under-doped (UD), optimally-doped (OP) and over-doped (OD) $Bi_2Sr_2CaCu_2O_{8+\delta}$ (Bi2212) superconductors with two $CuO_2$ planes in a unit cell.

Figure 1 shows the results of temperature versus in-plane resistance for the UD sample with $T_c$=74 K (Fig.1a), the OP sample with $T_c$=91 K (Fig.1b) and the OD sample with $T_c$=82 K (Fig.1c) at different pressures. It is found that the onset $T_c$ of these samples exhibits the same high-pressure behavior: a slight increase initially and then a monotonous decrease upon elevating pressure until not detectable. Subsequently, an unexpected insulating-like state presents at a pressure ($P_i$) of 34.3 GPa for the UD-doped sample, 39.9 GPa for the OP sample and 42.2 GPa for the OD sample, respectively. And the insulating-like behavior becomes pronounced when the pressure is higher than the $P_i$, (Fig.1a-1c). It is a grand surprise because naively one expects that by applying pressure the bandwidth should increase and thereby the system should become more metallic, but instead it becomes insulating-like. We repeated the measurements on new samples and found the results are reproducible [see Supplementary Information].

To investigate whether the insulating behavior observed is due to pressure-induced cracks in the material, we performed the high-pressure resistance and *ac* susceptibility experiments in the process of releasing pressure for the under-doped Bi2212 sample, and found a reversible transition between the superconducting state and the insulating-like state (see Supplementary Information). These results rule out the possibility that the transition from a superconducting state to an insulating-like state is caused by cracks.

The combined high-pressure measurements of *ac* susceptibility and in-plane resistance were performed for the above three kinds of samples. As shown in Fig.2, the superconducting transitions of the samples detected by the *ac* susceptibility can be clearly identified by the onset signal of the deviation from the almost constant background on the high-temperature side (see the blue plots) and the plunge of the resistance to zero (see the red plots). Upon compression to 34.3 GPa for the UD sample, 39.9 GPa for the OP sample and 42.2 for the OD sample, the samples show an insulating-like behavior (see the red plots in Fig.2d, 2h and 2l) and no diamagnetic signal is captured by the *ac* susceptibility measurements (see the blue plots in Fig.2d, 2h and 2l). These results manifest that the pressure induces a quantum phase transition from a superconducting state to an insulating-like state in all these superconductors.

We summarize the experimental results in the normalized pressure-$T_c$ phase diagram in the left panel of Fig.3, which is established on the basis of the pressure-$T_c$ phase diagrams of the UD, OP and OD Bi2212 samples (the right panels of Fig.3). The phase diagram for the three kinds of samples shows two distinct regions: the superconducting state (SC) and the insulating-like state (I), and demonstrates a universal quantum phase transition from the superconducting to the insulating-like states. It is seen that $T_c$ displays a slight increase initially within a small pressure range, and then a continuous decrease with elevating pressure until fully suppressed at a critical pressure ($P_c$, the determination of $P_c$ can be found in the caption of Figure 3), above which an insulating-like state emerges, as shown in the left panel of Fig.3 (the detail of the normalizing analysis can be found in the Supplementary information).

In order to know whether the quantum phase transition discovered in this study is a common phenomenon beyond the Bi2212 superconductors investigated, we conducted the same measurements on the $Bi_2Sr_{1.63}La_{0.37}CuO_{6+\delta}$ (Bi2201) superconductor with one $CuO_2$ plane and the $Bi_2Sr_2Ca_2Cu_3O_{10+\delta}$ (Bi2223) superconductor with three $CuO_2$ planes in a unit cell. The same phenomenon is also found in these superconductors (see Supplementary Information), indicating that the observed quantum phase transition is universal in these bismuth-bearing cuprate superconductors, regardless of the doping level and the number of $CuO_2$ planes in a unit cell.

These results impact our knowledge about the cuprate superconductors that, after the superconducting state is destroyed, the sample should show a well-behaved metallic state because pressure generally increases the bandwidth. To clarify the possible origin that leads to the destruction of the superconducting state and the emergence of the insulating-like state under pressure, we carried out more experiments.

First, we conducted the high-pressure synchrotron X-ray diffraction measurements at 50 K for the OD sample on beamline 4W2 at the Beijing Synchrotron Radiation Facility. Our results indicated that there is no structural phase transition in the range of pressure up to 43.1 GPa, except that the volume of the lattice is apparently compressed (see Supplementary Information). These results ruled out the possibility that the quantum phase transition from superconducting to insulating-like states connects with a pressure-induced structural phase transition.

Second, we measured the magnetoresistance (*MR*) at 4 K for the compressed UD, OP and OD samples that host the insulating-like state. The magnetic field was applied perpendicular to the *ab*-plane of these samples. As shown in Fig.4a-c, the *MR* of all the samples exhibits a positive effect, the in-plane resistance increases upon elevating magnetic field. Considering that the *MR* is very weak (~1%) and the appearance of the insulating-like state is close to the superconducting-insulating transition, we presume that the origin of the positive *MR* may be related to the superconducting fluctuation.

Third, we performed the high-pressure Hall coefficient ($R_H$) measurements for the OD sample (Fig.4d) and find that $R_H(P)$ decreases remarkably with increasing pressure up to ~18 GPa. Because the Hall resistance versus magnetic field displays a linear behavior in the pressure range investigated (Supplementary Information), a typical feature of the single band, the decrease of $R_H(P)$ below 18 GPa ought to be associated with the enhancement of carrier density. However, $R_H$ remains almost unchanged for pressures ranging from ~18 GPa to ~35 GPa and then shows a slow increase from ~35 GPa to 48.3 GPa. No apparent change in $R_H(P)$ at $P_c$ =39.5GPa implies that the total density of charge carriers seems to remain in a steady state across the quantum criticality. The reproducible result is also obtained in the Bi2201 superconductor (see Supplemental Information).

It is noted that, unlike the usual insulator, the low-temperature resistance in the insulating-like state rises way too slowly to be exponential. We attempted to fit the low temperature resistance with exponential dependence and power law, but they fail [see Supplementary Information]. Slow rises of the kind have been found in the low temperature orthorhombic $La_{2-x}Sr_xCuO_4$, $YiBa_2Cu_3O_{7-\delta}$ cuprates and $La_{1-x}M_xOBiS_2$, which are perceived as quite mysterious [32-35].

There is in fact no precedence anywhere else for such a transition from a superconducting state to an insulating-like state without a coincident structural phase transition. Therefore, some questions are raised naturally: why do the itinerant superconducting electrons become localized after the quantum phase transition, and what is the exotic pathway that results in the quantum phase transition? If considering the buckling of the $CuO_2$ planes due to the asymmetric lattice structure around the planes and nonuniform deformation from the doped atoms, when pressure shrinks the lattice constant of the planes, the plane buckling should be enhanced, which may help to develop an exotic band structure with a reduction of the bandwidth [36] and eventually result in the emergence of the insulating-like state. All the above are the attractive issues in searching for the new physics behind the pressure-induced quantum phase transition from a superconducting state to insulating-like state instead of to a metallic state, which deserves further investigation with other advanced experimental probes and sophisticated theoretical studies.

**Acknowledgements**

The work in China was supported by the National Key Research and Development Program of China (Grant No. 2017YFA0302900, 2016YFA0300300 and 2017YFA0303103), the NSF of China (Grants No. U2032214, 11888101 and 12004419) and the Strategic Priority Research Program (B) of the Chinese Academy of Sciences (Grant No. XDB25000000). We thank the support from the Users with Excellence Program of Hefei Science Center CAS (2020HSC-UE015). Part of the work is supported by the Synergic Extreme Condition User System. J. G. is grateful for support from the Youth Innovation Promotion Association of the CAS (2019008). The work in BNL was supported by the US Department of Energy, office of Basic Energy Sciences (contract No. desc0012704).

These authors with star (*) contributed equally to this work.

Correspondence and requests for materials should be addressed to L.S.(llsun@iphy.ac.cn).

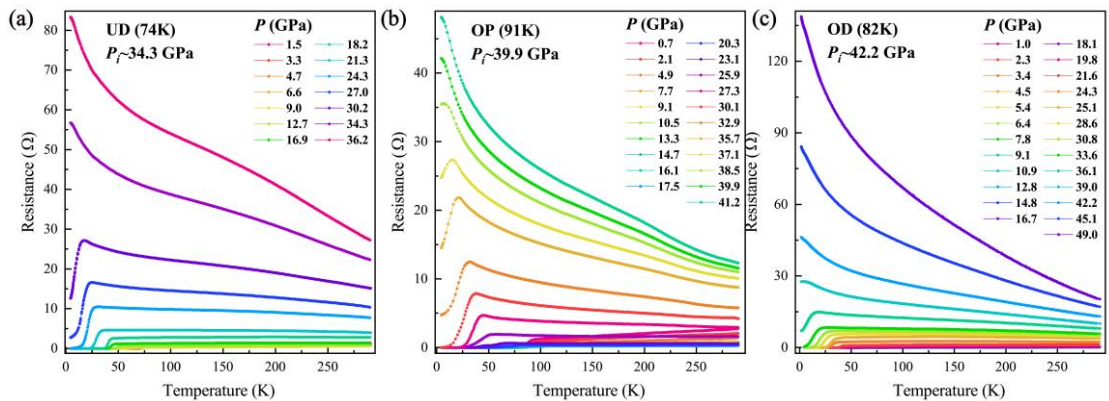

**Figure 1 Temperature dependence of in-plane resistance for Bi$_2$Sr$_2$CaCu$_2$O$_{8+\delta}$ (Bi2212) at different pressures: (a)** for the under-doped (UD) superconductor with superconducting transition temperature ($T_c$) about 74 K in the pressure range of 1.5 GPa – 36.2 GPa; **(b)** for the optimally-doped (OP) sample with $T_c$ about 91 K in the pressure

range of 0.7 GPa – 41.2 GPa; (**c**) for the over-doped (OD) sample with $T_c$ about 82 K in the pressure range of 1.0 GPa - 49 GPa. The three kinds of samples display the same behavior of an insulating-like state above the pressure ($P_i$).

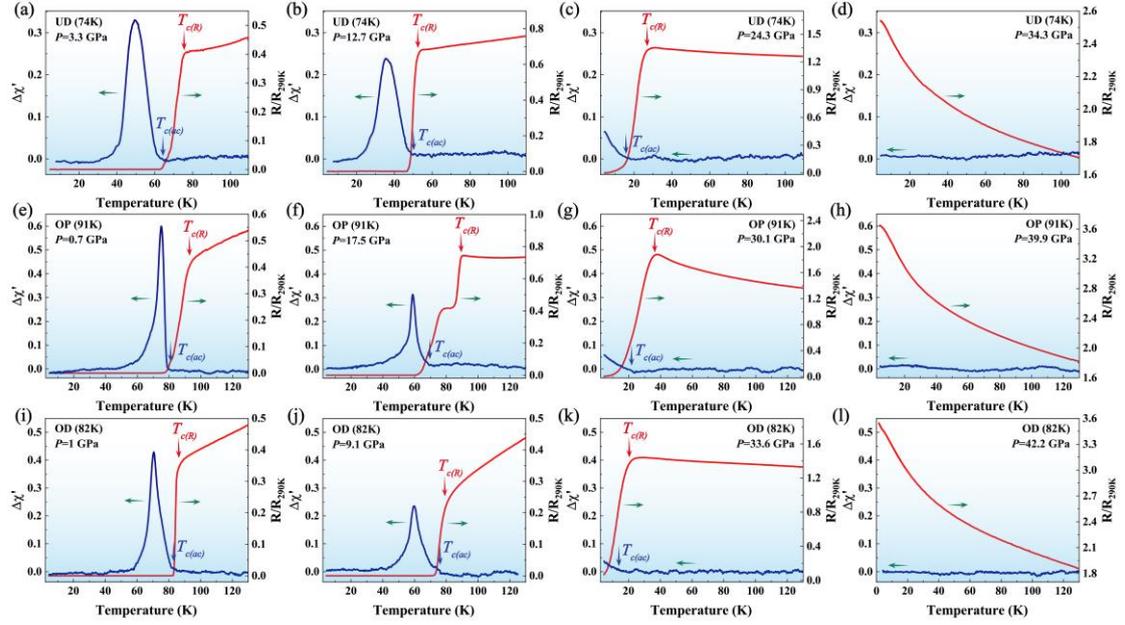

**Figure 2**. **In-plane resistance (*R*) and *ac* susceptibility ($\Delta\chi'$) as a function of temperature (*T*) for the Bi$_2$Sr$_2$CaCu$_2$O$_{8+\delta}$ superconductors at different pressures:** (**a**)-(**d**) for the under-doped (UD) superconductor; (**e**)-(**h**) for the optimally-doped (OP) superconductor; (**i**)-(**l**) for the over-doped (OD) superconductor. The blue lines in the figures are the data of $\Delta\chi'(T)$, while the red lines are the data of *R(T)*. The red and blue arrows indicate the temperatures of the onset superconducting transition detected by resistance and *ac* susceptibility measurements, respectively.

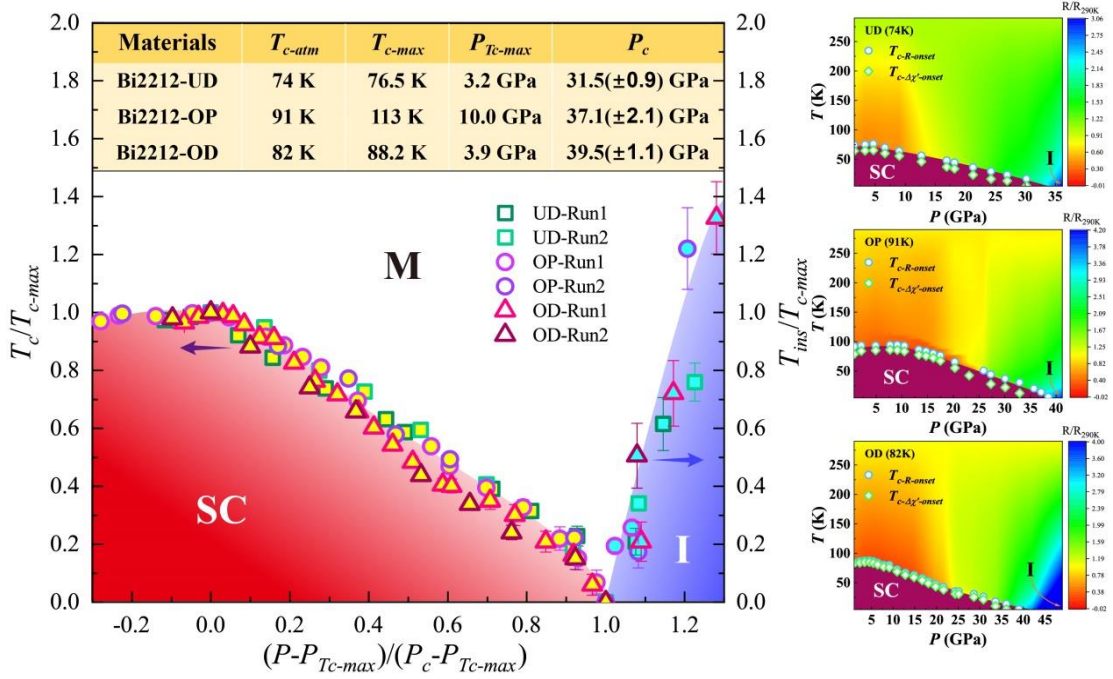

**Figure 3 Pressure-$T_c$ phase diagrams for Bi$_2$Sr$_2$CaCu$_2$O$_{8+\delta}$ superconductors.** The Right panels are the phase diagrams established by the experimental results from the under-doped (UD), optimally-doped (OP) and over-doped (OD) samples, together with the mapping information of temperature and pressure dependent $R$ (shown in color scale). The left panel is a normalized phase diagram that is built on the basis of the experimental phase diagrams (the right panels). $P_{Tc\text{-}max}$ and $P_c$ stand for the critical pressures where $T_c$ reaches the maximum and the zero, respectively. In the normalizing analysis, we define the pressure as $P_{Tc\text{-}max}$ when $(P-P_{Tc\text{-}max})/(P_c-P_{Tc\text{-}max}) = 0$, and the pressure as $P_c$ when $(P-P_{Tc\text{-}max})/(P_c-P_{Tc\text{-}max}) = 1$. The results of the normalizing analysis for $T_c/T_{c\text{-}max}$ versus $(P-P_{Tc\text{-}max})/(P_c-P_{Tc\text{-}max})$ and $T_{ins}/T_{c\text{-}max}$ versus $(P-P_{Tc\text{-}max})/(P_c-P_{Tc\text{-}max})$ show that the three kinds of samples display a universal quantum phase transition from the superconducting state to an insulating-like state. SC, M and I stand for superconducting state, metallic state and insulating-like state, respectively. The region of the M phase is determined by the critical value of $R/R_{290K}$ where the quantum phase

transition occurs. For example, when $R/R_{290K}$ is greater than 3, the over-doped sample is in the insulating-like state while, $R/R_{290K}$ is less than 3, the sample is in the metallic state (see right bottom panel). $T_{c\text{-}R\text{-}onset}$ and $T_{c\text{-}\Delta\chi'\text{-}onset}$ denote the onset temperatures of the superconducting transition detected by the resistance and *ac* susceptibility measurements, respectively. $T_{c\text{-}max}$ and $T_{ins}$ are the maximum value of $T_c$ and the characteristic temperature of the insulating-like transition (the method of determining the $T_{ins}$ can be found in the Supplementary information), respectively. The $P_c$ value is determined by the average pressure of two experimental runs $[P_c = (P'_{c\text{-}run1} + P'_{c\text{-}run2})/2]$, in which $P'_c$ of each experimental run is determined by the highest experimental pressure where the superconducting transition can still be observed and the lowest experimental pressure where the insulating-like state appears. The error bar of $P_c$ is the difference between $P'_{c\text{-}run1}$ and $P'_{c\text{-}run2}$.

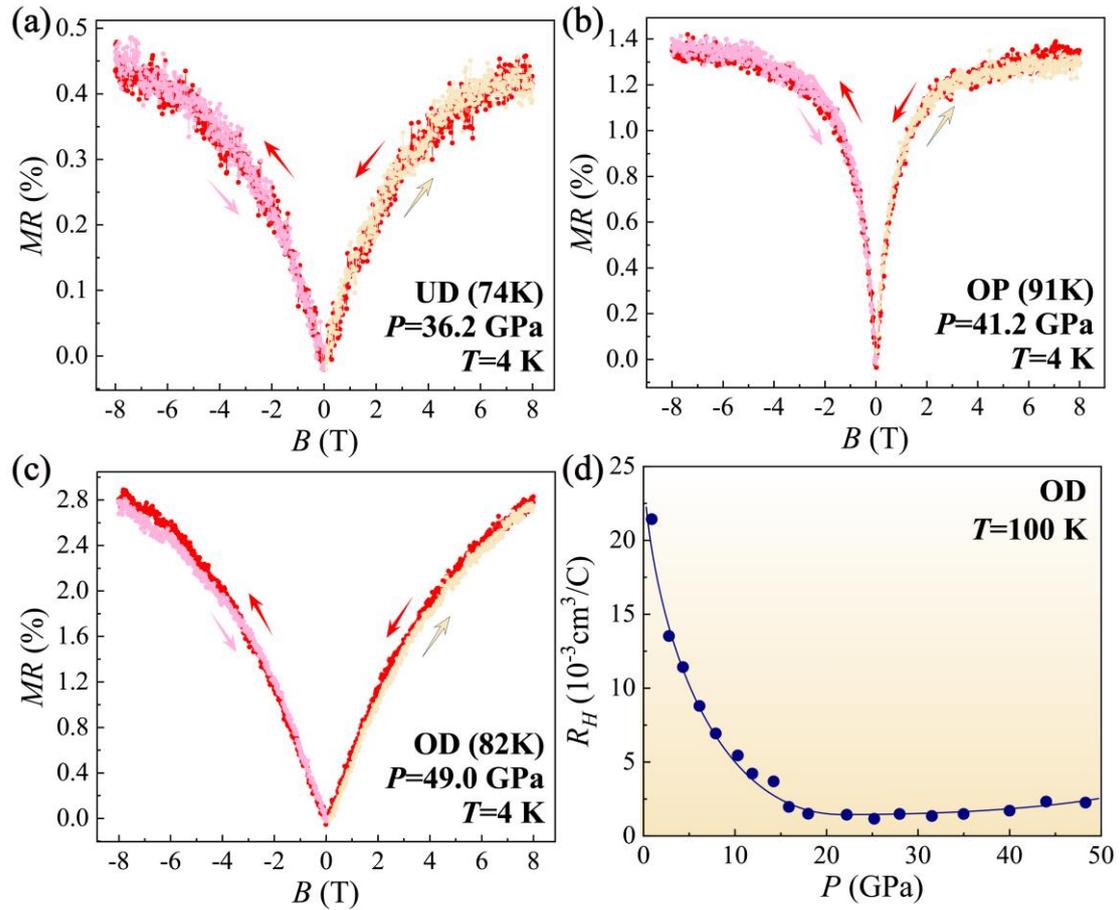

**Figure 4 Magnetoresistance (*MR*) as a function of magnetic field (*B*) for the UD, OP and OD Bi$_2$Sr$_2$CaCu$_2$O$_{8+\delta}$ superconductors when they enter into an insulating-like state, and Hall coefficient information of the OP sample under pressure. (a)-(b)** Plots of *MR* versus *B* for the UD, OP and OD samples measured at 4 K at 36.2 GPa, 41.2 GPa and 49 GPa, respectively. It is seen that all of them exhibit a positive magnetic effect. The red and beige arrows represent the directions of increasing and decreasing magnetic field. **(d)** Pressure dependence of Hall coefficient (*R$_H$*) for the OD superconductor measured at 100 K.